\documentclass[12pt,a4paper]{article}
\usepackage{epsfig}
\begin{document}
\title{Pair production of the heavy leptons in future high energy linear $e^{+}e^{-}$
colliders}
\author{Chong-Xing Yue, Nan Zhang,  Shi-Hai Zhu\\
{\small Department of Physics, Liaoning  Normal University, Dalian
116029, P. R. China}
\\}
\date{\today}

\maketitle
\begin{abstract}
The littlest Higgs model with T-parity predicts the existence of the
T-odd particles, which can only be produced in pair. We consider
pair production of the T-odd leptons in future high energy linear
$e^{+}e^{-}$ collider ($ILC$). Our numerical results show that, as
long as the T-odd leptons are not too heavy, they can be copiously
produced and their possible signals might be detected via the
processes $e^{+}e^{-}\rightarrow \overline{L}_{i}L_{j}$ in future
$ILC$ experiments.

\vspace{1cm}

\end{abstract}

\newpage
\noindent{\bf 1. Introduction}

Many popular specific models beyond the standard model $(SM)$
predict the existence of the heavy leptons. It is well known that,
so far, a clear signal of such new fermions has not been found at
high energy collider experiments. However, the experimental lower
bounds for the heavy lepton mass was found to be $44GeV$ by
$OPAL$[1], $46GeV$ by $HLEPH$[2], and $90GeV$ by $H1$
Collaborations[3]. This means that, if this kind of new particles
indeed exist, they should be detected in future high energy collider
experiments. Any signal for such kind of fermions in future high
energy experiments will play an important role in discovery of the
new physics beyond the $SM$.

Little Higgs theory[4] is proposed as an interesting solution to so
called hierarchy problem of the $SM$ and can be regarded as one of
the important candidates of the new physics beyond the $SM$. Among
of the little Higgs models, the Littlest Higgs $(LH)$ model[5] is
one of the simplest and phenomenologically viable models, which has
all essential features of the little Higgs models. However, the $LH$
model suffers from severe constraints from precision electroweak
measurement, which could only be satisfied by fine tuning the model
parameters[6]. To avoid this serious problem, a new discrete
symmetry (called T-parity) has been introduced, which forms called
$LHT$ model[7]. In this model, all dangerous tree level
contributions to low energy electroweak observables are forbidden by
T-parity and hence the corrections to observables are loop
suppressed. The $LHT$ model is one of the attractive little Higgs
models.

In order to implement T-parity in  fermion sector, one introduces
 three doublets of ``mirror quarks" and three doublets of ``mirror
 leptons", which have T-odd parity, transform vectorially under
 $SU(2)_{L}$ and can be given a large mass. A first study of the
 collider phenomenology of the $LHT$ model was given in Ref.[8]. The
 possible signals of the T-odd fermions (mirror fermions) have been
 studied in Refs.[9,10,11,12]. In this paper, we will focus our
 attention on the T-odd leptons and see whether the possible signals of
 the $LHT$ model can be
 detected in future high energy linear $e^{+}e^{-}$ collider $(ILC)$ experiments via
 the production which are related the T-odd leptons.

Studying production and decay of the new charged leptons at  high
energy collider experiments is of special interest. It will be
helpful to test the $SM$ flavor structure and new physics beyond the
$SM$. This fact has lead to many studies involving the new charged
leptons at $e^{+}e^{-}$ colliders[13], $ep$ colliders[14], and
hadron colliders[15].  Although there are lot of works on the new
charged leptons in the literature, it is need to be further studied
in the context of the $LHT$ model. There are several motivations to
perform this study. First, the $LHT$ model is one of the attractive
little Higgs models, which predicts the
 existence of the T-odd heavy charged leptons. However, in previous works
 on studying the phenomenology of the $LHT$ model, studies
 about the heavy charged leptons are very little. Second,  a pair of the T-odd
 leptons can be directly produced at the $CERN$ Large Hadron Collider
 $(LHC)$ through s-channel exchange of the $SM$ gauge
bosons. However, its production cross section is very small in most
of the parameter space of the $LHT$ model[9,12]. So far, a complete
study on pair production of the T-odd  charged leptons has not been
presented in the context of the $LHT$ model. Third, studying the
possible signals of the heavy charged leptons in future high energy
colliders can help the collider experiments to  test little Higgs
models and distinguish different new physics models. Thus, in this
paper, we will concentrate our attention on pair production of the
heavy charged leptons (T-odd) in future $ILC$ experiments.

In the present work, we study the dynamical properties for
production of the T-odd leptons and also for decay of the T-odd
leptons into presently known particles. We also discuss how the
signals can be clearly separated from the $SM$ backgrounds with a
great significance. After reviewing the $LHT$ model in section 2,
the production processes and signatures of the T-odd leptons are
studied in detail in section 3. Finally, our conclusions and simple
discussions are given in section 4.

\noindent{\bf 2. The $\emph{\textbf{LHT}}$ model}

In this section, we briefly review the essential features of the
$LHT$ model studied in Ref.[7], which are related our calculation.
Similar with the $LH$ model, the $LHT$ model is based on an
$SU(5)/SO(5)$ global symmetry breaking pattern. A subgroup
$[SU(2)\times U(1)]_{1}\times [SU(2)\times U(1)]_{2}$ of the $SU(5)$
global symmetry is gauged, and at the scale $f$ it is broken into
the $SM$ electroweak symmetry $SU(2)_{L}\times U(1)_{Y}$. T-parity
is an automorphism which exchanges the $[SU(2)\times U(1)]_{1}$ and
$[SU(2)\times U(1)]_{2}$ gauge symmetries. The T-even combinations
of the gauge fields are the $SM$ electroweak gauge bosons
$W_{\mu}^{a}$ and $B_{\mu}$. The T-odd combinations are T-parity
partners of the $SM$ electroweak gauge bosons.

After taking into account electroweak symmetry breaking $(EWSB)$, at
the order of $\nu^{2}/f^{2}$, the masses of the T-odd set of the
$SU(2)\times U(1)$ gauge bosons are given as:

\begin{equation}
M_{B_{H}}=\frac{g'f}{\sqrt{5}} [1-\frac{5\nu^{2}}{8f^{2}}],
\hspace{0.5cm}M_{Z_{H}} \approx
M_{W_{H}}=gf[1-\frac{\nu^{2}}{8f^{2}}].
\end{equation}
Where $\nu=246GeV$ is the electroweak scale and $f$ is the scale
parameter of the gauge symmetry breaking of the $LHT$ model. $g'$
and $g$ are the $SM$ $U(1)_{Y}$ and $SU(2)_{L}$ gauge coupling
constants, respectively. Because of the smallness of $g'$, the T-odd
gauge boson $B_{H}$ is the lightest T-odd particle, which can be
seen as an attractive dark matter candidate[16].

To avoid severe constraints and simultaneously implement T-parity,
it is need to double the $SM$ fermion doublet spectrum[7,8,17]. The
T-even combination is associated with the $SU(2)_{L}$ doublet, while
the T-odd combination is its T-parity partner. The masses of the
T-odd fermions can be written in a unified manner as:

\begin{equation}
M_{F_{i}}=\sqrt{2}k_{i}f,
\end{equation}
where $k_{i}$ are the eigenvalues of the mass matrix $k$ and their
values are generally dependent on the fermion species $i$.

The mirror fermions (T-odd quarks and T-odd leptons) have new flavor
violating interactions with the $SM$ fermions mediated by the new
gauge bosons $(B_{H},W_{H}^{\pm}$, or $Z_{H})$, which are
parameterized by four $CKM$-$like$ unitary mixing matrices, two for
mirror quarks and two for mirror leptons[11,12,18]:
\begin{equation}
V_{Hu},\hspace*{0.2cm}V_{Hd},\hspace*{0.2cm}V_{Hl},\hspace*{0.2cm}V_{H\nu}.
\end{equation}
They satisfy:
\begin{equation}
V_{Hu}^{+}V_{Hd}=V_{CKM},\hspace*{0.2cm}V_{H\nu}^{+}V_{Hl}=V_{PMNS}.
\end{equation}

Where the $CKM$ matrix $V_{CKM} $ is defined through flavor mixing
in the down-type quark sector, while the $PMNS$ matrix $V_{PMNS} $
is defined through neutrino mixing. Similar with Ref.[11], we will
set the Majorana phases of $V_{PMNS}$ to zero in our following
calculation. The matrix $V_{Hl}$ can give rise to the lepton flavor
violating processes.

The couplings of the T-odd leptons to other particles, which are
related our analysis, are summarized as[11]:

\begin{eqnarray}
Z\overline{L}_{i}L_{j}&:&\frac{ie}{S_{w}C_{w}}[-\frac{1}{2}+S_{w}^{2}]
\gamma^{\mu}\delta_{ij},
\hspace*{0.4cm}\gamma\overline{L}_{i}L_{j}:-ie\gamma^{\mu}\delta_{ij};
\\Z_{H}\overline{L}_{i}l_{j}&:&\frac{ie}{S_{w}}[-\frac{1}{2}+\frac{S_{w}^{2}}
{8(5C_{w}^{2}-S_{w}^{2})}\frac{\nu^{2}}{f^{2}}](V_{Hl})_{ij}\gamma^{\mu}P_{L};
\\B_{H}\overline{L}_{i}l_{j}&:&\frac{ie}{C_{w}}[\frac{1}{10}+\frac{5C_{w}^{2}}
{8(5C_{w}-S_{w}^{2})}\frac{\nu^{2}}{f^{2}}](V_{Hl})_{ij}\gamma^{\mu}P_{L}.
\end{eqnarray}
Where $S_{w}=\sin\theta_{w},\theta_{w}$ is the Weinberg angle.
$l_{i}$ and $L_{j}$ represent the three family leptons $e, \mu,$ or
 $\tau$ and their T-odd partners,
 respectively. $P_{L}=(1-\gamma_{5})/2$ is the left-handed projection
 operator.

From above discussions, we can see that the $LHT$ model provides a
new mechanism for lepton flavor violation $(LFV)$, which comes from
the flavor mixing in the mirror lepton sector. Thus, the $LHT$ model
can give significantly contributions to some $LFV$ processes, such
as $l_{i}\rightarrow l_{j}\gamma$, $l_{i}\rightarrow
l_{j}l_{k}l_{l}$, $\tau\rightarrow\mu\pi$ etc[19]. In the next
section, we will consider pair production of the T-odd leptons in
future $ILC$ experiments and further discuss their $LFV$ signatures.

\noindent{\bf 3. Pair production of the T-odd leptons at the
$\emph{\textbf{ILC}}$}

 In the $LHT$ model[7], T-parity explicitly forbids the tree level
 contributions coming from the new particles to the observables involving only the
 $SM$ particles and forbids the interactions that induce triplet vacuum expectation
 value $(VEV)$ contributions. The $SM$ particles are T-even, while the new particles
  are T-odd, except for the T-parity partner
 of the top quark. As a consequence, the electroweak precision measurement data
 allow for a relatively low value of the new particle mass scale
 $f\sim500GeV$ and the T-odd particles can only be produced in
 pairs. Pair production of the T-odd particles has been studied via
 $pp$[9,12], $e\gamma$ and $ep$ collisions[20].

 \vspace{-5.5cm}
 \begin{figure}[htb]
\begin{center}
\epsfig{file=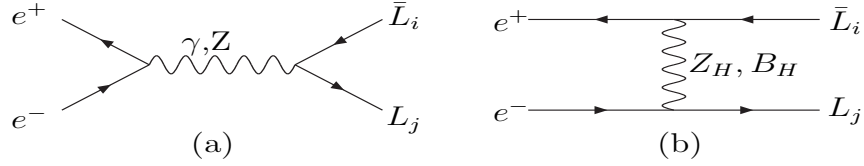,width=650pt,height=720pt} \vspace{-16.85cm}
 \caption{Feynman diagrams for pair production of the T-odd leptons at the $ILC$.}
  \label{ee}
\end{center}
\end{figure}

 From above discussions, we can see that pair production of the T-odd
 leptons at $ILC$ proceeds via the s-channel and t-channel Feynman
 diagrams as shown in Fig.1. The invariant scattering amplitude for
 the process $e^{+}(P_{1})e^{-}(P_{2})\rightarrow \bar{L}_{i}(P_{3})L_{j}(P_{4})$
 can be written as:
 \begin{eqnarray}
 iM&=&\frac{e^{2}}{K_{1}^{2}}\bar{v}(P_{1})\gamma^{\mu}u(P_{2})\bar{u}(P_{4})
 \gamma_{\mu}v(P_{3}) \nonumber\\
 &&+\frac{e^{2}}{4S_{w}^{2}C_{w}^{2}}
 \frac{1}{K_{1}^{2}-M_{Z}^{2}}[-\frac{1}{2}+S_{w}^{2}]\bar{v}(P_{1})(4S_{w}^{2}-1
 +\gamma_{5})\gamma^{\mu}
 u(P_{2})
 \bar{u}(P_{4})\gamma_{\mu}v(P_{3})\nonumber\\&&+\frac{a^{2}(V_{Hl})_{ei}(V_{Hl})_{ej}}{K_{2}^{2}-M_{Z_{H}}^{2}}
 \bar{v}(P_{1})\gamma^{\mu}P_{L}v(P_{3})
 \bar{u}(P_{4})
 \gamma_{\mu}P_{L}u(P_{2})\nonumber\\&&+\frac{b^{2}(V_{Hl})_{ei}(V_{Hl})_{ej}}{K_{2}^{2}-M_{Z_{H}}^{2}}\bar{v}
 (P_{1})\gamma^{\mu}P_{L}v(P_{3})
 \bar{u}(P_{4})\gamma_{\mu}
 P_{L}u(P_{2}),
 \end{eqnarray}
where
\begin{eqnarray}
&&\hspace*{-1.0cm}K_{1}^{2}=(P_{1}+P_{2})^{2},\hspace{4.0cm}
K_{2}^{2}=(P_{4}-P_{2})^{2};
\\&&\hspace*{-1.0cm}a=\frac{e}{S_{w}}[-\frac{1}{2}+
\frac{S_{w}^{2}}{8(5C_{w}^{2}-S_{w}^{2})}\frac{v^{2}}{f^{2}}],
\hspace*{0.3cm}b=\frac{e}{C_{w}}[-\frac{1}{10}+
\frac{5C_{w}^{2}}{8(5C_{w}^{2}-S_{w}^{2})}\frac{v^{2}}{f^{2}}].
\end{eqnarray}

From Eqs.(8), (9), and (10) we can see that, except the $SM$ input
parameters $\alpha_{e}=1/128.8$, $S_{w}^{2}=0.2315$, and
$M_{Z}=91.187GeV$[21], the production cross sections
$\sigma(\overline{L}_{i}L_{j})$ for the processes
$e^{+}e^{-}\rightarrow \overline{L}_{i}L_{j}$ are dependent on the
model dependent parameters $f,k($or $M_{L_{i}})$, $(V_{Hl})_{ei}$,
and $(V_{Hl})_{ej}$. The matrix elements $(V_{Hl})_{ij}$ can be
determined through $V_{Hl}=V_{H\nu}V_{PMNS}$. To avoid any
additional parameters introduced and to simply our calculation, we
take $V_{Hl}=V_{PMNS}$, which means that the T-odd leptons have no
impact on flavor violating observables in the neutrino sector. For
the matrix $V_{PMNS}$, we take the standard parameterization form
with parameters given by the neutrino experiments[22,23]. Ref.[19]
has shown that, for $V_{Hl}=V_{PMNS}$, to make the $\mu \rightarrow
e\gamma$ and $\mu^{-}\rightarrow e^{-}e^{+}e^{-}$ decay rates
consistent with the present experimental upper bounds, the spectrum
of the T-odd leptons must be quasi-degenerate. Thus, in our
numerical estimation, we will assume
$M_{L_{e}}=M_{L_{\mu}}=M_{L_{\tau}}=M_{L}$ and take the parameters
$f$ and $M_{L}$ as free parameters.

\begin{figure}[htb]
\begin{center}
\epsfig{file=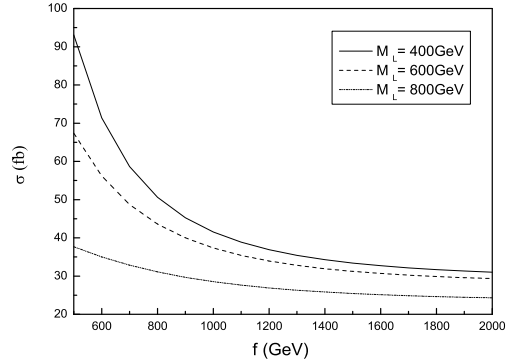,width=210pt,height=160pt} \vspace{-1.0cm}
 \caption{The production cross section $\sigma(\overline{L}_{\mu}L_{\mu}$)
  as a function of the scale parameter $f$ \hspace*{2.0cm}for $\sqrt{s}=2$TeV
  and three
   values of the T-odd lepton mass $M_{L}$.} \label{ee}
\end{center}
\end{figure}

Our numerical results are shown in Fig.2 and Fig.3, in which we plot
the production cross sections $\sigma(\bar{L}_{\mu}L_{\mu})$ and
$\sigma(\bar{L}_{e}L_{\mu})$ as functions of the scale parameter $f$
for the center-of-mass $\sqrt{s}=2TeV$ and three values of the T-odd
lepton mass $M_{L}$. Since the value of the matrix element
$(V_{PMNS})_{e\tau}$ is smaller than that of $(V_{PMNS})_{e\mu}$,
the production cross sections $\sigma(\bar{L}_{\tau}L_{\tau})$ and
$\sigma(\bar{L}_{e}L_{\tau})$[or $\sigma(\bar{L}_{\mu}L_{\tau})$]
are smaller than $\sigma(\bar{L}_{\mu}L_{\mu})$ and
$\sigma(\bar{L}_{e}L_{\mu})$, respectively. So, in Fig.2 and Fig.3,
we have not given the curves for the production cross sections
$\sigma(\bar{L}_{\tau}L_{\tau})$, $\sigma(\bar{L}_{e}L_{\tau})$, and
$\sigma(\bar{L}_{\mu}L_{\tau})$. Using the unitarity based $PDG$
parametrization and available data from oscillation experiments,
Refs.[22,23] have constructed the $PMNS$ matrix $V_{PMNS}$, in which
the values of the matrix elements $(V_{PMNS})_{e\mu}$ and
$(V_{PMNS})_{ee}$ are given in the ranges of 0.4871$\sim$ 0.6193 and
0.7575 $\sim$ 0.8819, respectively. To simply our calculation, we
have taken the values of $(V_{PMNS})_{e\mu}$ and $(V_{PMNS})_{ee}$
as 0.55 and 0.82 in Fig.3 and Fig.2, respectively. From Fig.(2) and
Fig.(3), we can see that the values of the production cross sections
increase as the scale parameter $f$ decreasing and as the T-odd
lepton mass $M_{L}$ decreasing. For $M_{L}=400GeV$ and $500GeV\leq
f\leq 2000GeV$, the values of $\sigma(\bar{L}_{\mu}L_{\mu})$ and
$\sigma(\bar{L}_{e}L_{\mu})$ are in the ranges of $93.1fb\sim31fb$
and $171.5fb\sim33.5fb$, respectively. While for $M_{L}=800GeV$ and
$500GeV\leq f\leq 2000GeV$, their values are in the ranges of
$37.7fb\sim24.3fb$ and $55.2fb\sim 25.8fb$, respectively. If we
assume that the $ILC$ experiment with $\sqrt{s}=2TeV$ has a yearly
integrated luminosity of $\pounds=100fb^{-1}$ and assume
$M_{L}<900GeV$ and $f\leq2TeV$, then there will be several hundreds
up to thousands of $\bar{L}_{\mu}L_{\mu}$ or $\bar{L}_{e}L_{\mu}$
events to be generated per year.

\begin{figure}[htb]
\begin{center}
\epsfig{file=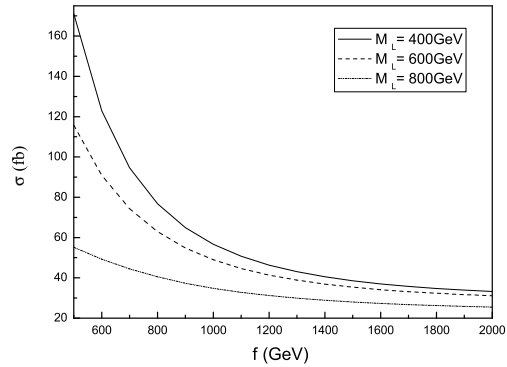,width=210pt,height=160pt} \vspace{-1.0cm}
 \caption{Same as Fig.2 but for the production cross section $\sigma(\bar{L}_{e}L_{\mu})$ }
  \label{ee}
\end{center}
\end{figure}

 The new gauge boson $B_{H}$ is the lightest T-odd particle, which can be
seen as an attractive dark matter candidate[16].  The T-odd lepton
$L_{i}$ mainly decays to $B_{H}l_{i}(l_{i}=e$, $\mu$ or $\tau)$ and
there is $Br(L_{i}\rightarrow B_{H}l_{i})$$\approx$100\%[9,12,23].
In this case, the signature of the process $e^{+}e^{-}\rightarrow
\bar{L}_{\mu}L_{\mu}$  is the opposite-sign same-flavor leptons
$\bar{\mu}\mu$ plus large missing energy i.e.
$\bar{\mu}\mu+B_{H}B_{H}$. The large transverse missing energy can
be used to distinguish the signal events from the $SM$ signal events
generated by the process
$e^{+}e^{-}\rightarrow\gamma/Z\rightarrow\bar{\mu}\mu$, which can
not be considered as background. The intrinsic $SM$ backgrounds come
from the processes
$e^{+}e^{-}\rightarrow\mu\bar{\mu}Z\rightarrow\mu\bar{\mu}\nu\bar{\nu}$
and $e^{+}e^{-}\rightarrow
W^{+}W^{-}\rightarrow\mu\bar{\mu}\nu_{\mu}\bar{\nu_{\mu}}$. However,
at the $ILC$ experiment with $\sqrt{s}=2TeV$, the cross section of
the former process is about $2fb$ while the one of the latter
process is about $11.2fb$, which is smaller than the cross section
of the process $e^{+}e^{-}\rightarrow \bar{L_{\mu}}L_{\mu}$ in most
of the parameter space of the $LHT$ model. Thus, the signal event
$\bar{\mu}\mu+B_{H}B_{H}$ should be easily separated from the $SM$
background with a great significance. We expect that, as long as it
is not too heavy, the T-odd lepton should be detected via the
process $e^{+}e^{-}\rightarrow \bar{L}_{\mu}L_{\mu}$ in future $ILC$
experiments.

\begin{figure}[htb]
\begin{center}
\epsfig{file=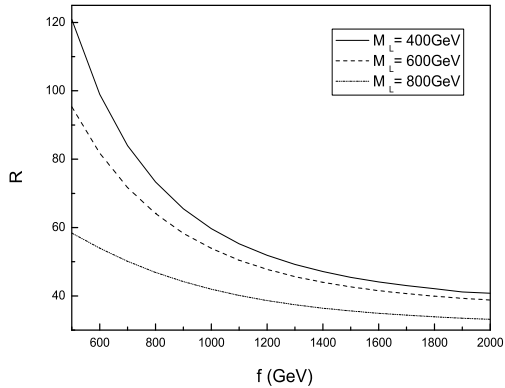,width=210pt,height=170pt} \vspace{-1.0cm}
 \caption{The statistical significance $R=S/\sqrt{B}$ as a function of
 the scale parameter $f$ \hspace*{1.85cm}for three values of the T-odd
 lepton mass $M_{L}$.} \label{ee}
\end{center}
\end{figure}

The $LFV$ process $e^{+}e^{-}\rightarrow \bar{L}_{e}L_{\mu}$ can
give rise to the signal events with opposite-sign and
different-flavor leptons and large missing energy $(\bar{e} \mu
+\not\!\!E_{T})$, i.e. $e^{+}e^{-}\rightarrow
\bar{L}_{e}L_{\mu}\rightarrow \bar{e}\mu B_{H}B_{H}$. Although the
$LFV$ signal is quite spectacular, it is not free of the $SM$
background. The leading $SM$ backgrounds of the signal event
$\bar{e} \mu +\not\!\!E_{T}$ mainly come from the $WW$ pair
production process $e^{+}e^{-}\rightarrow \ W^{+}W^{-}\rightarrow
\bar{e}\mu \nu_{e}\overline{\nu_{\mu}}$. To see whether the $LFV$
signals of the T-odd leptons can be detected in future $ILC$
experiments, we further calculate the ratio of the signal over the
square root of the background $R=S/\sqrt{B}$, which is called
statistical significance. Our numerical results are shown in Fig.4,
in which we have taken the integrated luminosity
$\pounds=100fb^{-1}$ for $\sqrt{s}=2TeV$ and the branching ratios
$Br(W^{+}\rightarrow\bar{e}\nu_{e})=(10.66\pm0.17)\%$  and
$Br(W^{-}\rightarrow\ \mu\bar{\nu}_{\mu})=(10.60\pm0.15)\%$[21].
From Fig.4 one can see that, in most of the parameter space of the
$LHT$ model, the value of the statistical significance $R$ is larger
than $33$. Furthermore, if we apply appropriate cuts on the $SM$
backgrounds, the value of the ratio $R$ can be clearly improved. For
example, Ref.[24] has shown that appropriate kinematical cuts can
strongly reduce the $WW$ background. Thus, the possible signals of
the T-odd leptons should be easy detected via the $LFV$ process
$e^{+}e^{-}\rightarrow \bar{L}_{e}L_{\mu}$ in future $ILC$
experiments.

\noindent{\bf 4. Conclusions and discussions }

The $LHT$ model is one of the attractive little Higgs models, which
provides a possible dark mater candidate. To simultaneously
implement T-parity, the $LHT$ model introduces new mirror fermions
(T-odd quarks and T-odd leptons). The flavor mixing in the mirror
fermion sector gives rise to a new source of flavor violation, which
might generate significantly contributions to some flavor violation
processes.

The T-odd leptons can only be produced through weak processes and
their production cross sections are generally small at the $LHC$.
So, in this paper, we study pair production of the T-odd lepton in
future $ILC$ experiments. Our numerical results show that, as long
as the T-odd leptons are not too heavy, they can be copiously
produced in pairs. For example, for $500GeV\leq f\leq 2000GeV$ and
$M_{L}=600$GeV, the production cross section for the process
$e^{+}e^{-}\rightarrow \bar{L}_{\mu}L_{\mu}$ is in the range of
$67.4fb\sim 29.3fb$. Furthermore, the pair production process
$e^{+}e^{-}\rightarrow \bar{L}_{\mu}L_{\mu}$ can generate nice
signal event $\mu\bar{\mu} +\not\!\!E_{T}$, which might be easily
separated from the $SM$ background with a great significance.

The T-odd leptons can also be produced in pair via the $LFV$
processes $e^{+}e^{-}\rightarrow \bar{L}_{i}L_{j}(i\neq j)$. Except
the free parameters $f$ and $M_{L}$, their production cross sections
are dependent on the $PMNS$ matrix elements $(V_{PMNS})_{ei}$ and
$(V_{PMNS})_{ej}$. Considering the bounds of the neutrino
oscillation experiment data on these matrix elements, we calculate
the production cross section of the $LFV$ process
$e^{+}e^{-}\rightarrow \bar{L}_{e}L_{\mu}$. We find that its value
can be significantly large in most of the parameter space of the
$LHT$ model. The $SM$ backgrounds of this process mainly come from
the $SM$ process $e^{+}e^{-}\rightarrow W^{+}W^{-}$. Even if no cuts
are applied and the electron beam and the position beam are not
polarized, the value of the ratio $R$ can be larger than 33 in most
of the parameter space.

In conclusion, we have considered pair production of the T-odd
leptons and discussed the possible of detecting these new particles
in future $ILC$ experiments. We find that, as long as the T-odd
leptons are not too heavy, they can be copiously produced in pairs
via the processes $e^{+}e^{-}\rightarrow \bar{L}_{i}L_{j}$ and their
signatures might be observed in future $ILC$ experiments. Thus, we
expect that the future $ILC$ experiments can be seen as an ideal
tool to detect the T-odd leptons predicted by the $LHT$ model. Even
if we can not observe the signals in future $ILC$ experiments, at
least, we can obtain the bounds on the free parameters of the $LHT$
model.

The $LHT$ model might give significantly contributions to some $LFV$
processes, such as $l_{i}\rightarrow l_{j}\gamma$, $l_{i}\rightarrow
l_{j}l_{k}l_{l}$, $\tau\rightarrow\mu\pi$ etc. The present
experimental upper bounds of branching ratios $Br(\mu\rightarrow
e\gamma)$ and $Br(\mu^{-}\rightarrow e^{-}e^{+}e^{-})$ can give
severe constraints on the free parameters of the $LHT$ model[19].
Considering these constraints, we have assumed
$M_{L_{e}}=M_{L_{\mu}}=M_{L}$ for $V_{Hl}=V_{PMNS}$ in our numerical
estimation. From our numerical results, we can see that the values
of the cross section $\sigma(\overline{L}_{e}L_{\mu})$ and
$\sigma(\overline{L}_{\mu}L_{\mu})$ increase as the scale parameter
$f$ decreasing, which is similar with that for the branching ratios
$Br(\mu\rightarrow e\gamma)$ and $Br(\mu^{-}\rightarrow
e^{-}e^{+}e^{-})$. However, even for $M_{L}=400GeV$ and
$f\leq2000GeV$, the values of $\sigma(\overline{L}_{\mu}L_{\mu})$
and $\sigma(\overline{L}_{e}L_{\mu})$ are larger than $31fb$ and
$33fb$, respectively. Thus, we can say that the strong constraints
on the $LHT$ model, which come from the $LFV$ processes
$\mu\rightarrow e\gamma$ and $\mu^{-}\rightarrow e^{-}e^{+}e^{-}$,
do not strongly change our conclusions about production of the T-odd
leptons in future $ILC$ experiments.

\vspace{1.0cm}

\noindent{\bf Acknowledgments}

This work was supported in part by Program for New Century Excellent
Talents in University(NCET-04-0290), the National Natural Science
Foundation of China under the Grants No.10475037 and 10675057.

\end{document}